\documentclass[11pt]{article}
\usepackage{graphicx,amsmath,amsthm,mathrsfs,amssymb}
\usepackage[usenames]{color}
\usepackage{ulem}

\newcommand{\ee}{\end{equation}}

\newcommand{\reff}[1]{(\ref{#1})}
\newcommand{\beq}{\begin{equation}}
\newcommand{\eeq}[1]{\label{#1}\end{equation}}
\newcommand{\beqa}{\begin{eqnarray}}
\newcommand{\eea}{\end{eqnarray}}
\newcommand{\eeqa}[1]{\label{#1}\end{eqnarray}}
\newcommand{\beg}{\begin{equation*}}
\newcommand{\eeg}{\end{equation*}}

\newcommand{\eq}{\!=\!}

\newcommand{\m}{\!-\!}

\newcommand{\bsplit}{\begin{split}}
\newcommand{\esplit}{\end{split}}

\begin{document}
\begin{titlepage}
\title{Black hole free energy during charged collapse:\\ a numerical study}
\author{\thanks{hbeauchesne10@ubishops.ca} Hugues Beauchesne and \thanks{aedery@ubishops.ca} Ariel Edery \\\\{\small\it Physics Department, Bishop's University}\\
{\small\it 2600 College Street, Sherbrooke, Qu\'{e}bec, Canada
J1M~0C8}}
\date{}
\maketitle
\begin{abstract}
We perform a numerical investigation of the thermodynamics during the collapse of a charged (complex) scalar field to a Reissner-Nordstr\"om (RN) black hole in isotropic coordinates. Numerical work on gravitational collapse in isotropic coordinates has recently shown that the negative of the total Lagrangian approaches the Helmholtz free energy $F\eq E\m TS$ of a Schwarzschild black hole at late times of the collapse (where $E$ is the black hole mass, $T$ the temperature and $S$ the entropy). The relevant thermodynamic potential for the RN black hole is the Gibbs free energy $G\eq E \m TS\m \Phi_H Q$ where $Q$ is the charge and $\Phi_H$ the electrostatic potential at the outer horizon. In charged collapse, there is a large outgoing matter wave which prevents the exterior from settling quickly to a static state. However, the interior region is not affected significantly by the wave. We find numerically that the interior contribution to the Gibbs free energy is entirely gravitational and accumulates in a thin shell just inside the horizon. The entropy is gravitational in origin and one observes dynamically that it resides on the horizon. We also compare the numerical value of the interior Lagrangian to the expected analytical value of the interior Gibbs free energy for different initial states and we find that they agree to within $10\!-\!13\%$. The two values are approaching each other so that their difference decreases with more evolution time.
 
\end{abstract}
\end{titlepage}
\setcounter{page}{1}
\section{Introduction}

Recent numerical studies of gravitational collapse in isotropic coordinates, both in $4 +1$ dimensions (5D) \cite{Khlebnikov, C-E} and 4D \cite{C-E}, have provided numerical evidence that the negative of the total Lagrangian at late times of the collapse process approaches the Helmholtz free energy $E\m TS$ of a Schwarzschild black hole, where $E$, $T$ and $S$ are the mass, temperature and entropy of the black hole. For a stationary black hole, $T\eq \hbar \kappa/2\pi$ and $S\eq A/4\hbar$ where $\kappa$ is the surface gravity at the horizon and $A$ is the horizon area. Though $T$ and $S$ both contain $\hbar$, their product $TS$ does not and this opens the possibility for a classical investigation of black hole thermodynamics via the free energy \cite{Khlebnikov}. There is in fact an argument based on the Euclidean action that relates the negative of the total Lagrangian to the free energy of a stationary black hole \cite{Khlebnikov,Hawking2}. This association can be tested numerically during gravitational collapse by tracking the Lagrangian and comparing its numerical value at late times to the expected free energy from standard black hole thermodynamics. This was carried out numerically for the first time for the collapse of a 5D Yang-Mills instanton \cite{Khlebnikov} to a Schwarzschild black hole in isotropic coordinates. A numerical study in isotropic coordinates was then carried out for the collapse of a 4D and 5D massless scalar field to a Schwarzschild black hole \cite{C-E}. These works constituted a classical numerical study of black hole thermodynamics; quantum mechanics does not enter the picture and Hawking radiation is not observed. Prior studies of thermodynamics during gravitational collapse have consisted mainly of analytical or semi-analytical work on black hole entropy and Hawking radiation during shell or dust collapse \cite{Gao}-\cite{Vaz1} and the approaches have been quantum mechanical or semi-classical.  

For the Reissner-Nordstr\"om (RN) black hole, the relevant thermodynamic potential is the Gibbs free energy $G\eq E\m TS\m \Phi_H\,Q$ where $\Phi_H$ is the electrostatic potential at the horizon and $Q$ is the charge of the black hole \cite{Hawking2, Kiritsis}. In this paper, we investigate numerically in isotropic coordinates the thermodynamics during the collapse of a charged scalar field to a RN black hole. Among other things, we study the association between the total Lagrangian and the Gibbs free energy. Charged collapse leads to a large outgoing matter wave and the exterior takes a long time to settle to a static state. However, the interior spacetime (inside the outer horizon) is hardly affected by this wave and settles more quickly than the exterior. One can show analytically that the interior Gibbs free energy $G_{int}$ is equal to $-TS$. The interior contribution of the negative of the total Lagrangian can be obtained numerically and then compared to $G_{int}$. At late times of the collapse, we find that the matter Lagrangian tends towards zero in the interior; the interior Lagrangian stems entirely from the gravitational sector. It is negative and accumulates just inside the horizon where the spacetime is not static. In short, the entropy of the charged black hole is gravitational in origin and stems from the dynamical interior near the horizon. That the entropy stems from the interior is in accord with analytical work on the canonical quantization of the RN black hole \cite{Vaz2} where the entropy was obtained via explicit counting of microstates in the dynamical interior (see \cite{Vaz1} for the uncharged case). Our numerical results for the negative of the gravitational Lagrangian in the interior agrees with the analytical value of $G_{int}$ to within $10\!-\!13\%$ depending on the profile of the initial state. There are sharp changes in the gradients of the metric and matter functions in the near-horizon region and this places limits on how far in time one can evolve before the usual monitors of the code, such as the ADM mass, begin to deviate from their conserved values. The numerical graph representing the Lagrangian and the analytical graph representing the free energy approach each other with time and it is clear that given more time evolution the difference between them would continue to decrease. Of course, to increase the time evolution one needs to increase the resolution and one reaches a limit where the computing time is no longer practical. As a consistency check, we also implement a procedure for prolonging the evolution time in the exterior region. The outgoing matter wave disperses and the metric in the exterior is observed to approach the RN static exterior metric. Unlike Schwarzschild, the matter Lagrangian now makes a contribution to the exterior due to the presence of a static electric field.   

The coordinate time $t$ in isotropic coordinates coincides with the time measured by an asymptotic observer at rest. The viewpoint of the static asymptotic observer is appropriate for studying black hole thermodynamics. The temperature of a stationary black hole is the temperature as seen by an observer at rest at spatial infinity \cite{Lindesay} and its entropy represents a measure of an external observer's ignorance of the internal configurations hidden behind the event horizon \cite{Bekenstein, E-C}. In particular, the association between the total Lagrangian and the free energy of a black hole that is found in numerical studies of collapse in isotropic coordinates will in general not hold in other coordinate systems. Unlike the action, the Lagrangian depends on the choice of time coordinate i.e. on the foliation of the spacetime. For example, one should not expect the Lagrangian in numerical studies of gravitational collapse in Painlev\'{e}-Gullstrand (PG) coordinates \cite{Kunstatter} to be associated with the free energy of the black hole. The reason is that in PG coordinates the coordinate time represents the proper time of freely-falling observers \cite{Poisson} not asymptotic observers. This is also true of numerical studies of charged collapse that have been carried out in ``Eddington-Finkelstein" \cite{Brady} and double-null coordinates \cite{Hod,Piran} as there is again no coordinate representing the proper time of an asymptotic observer in these cases.      

Our paper proceeds as follows. We first express the exterior RN metric in isotropic coordinates. We then derive the equations of motion in those coordinates: the wave equation governing the complex (charged) scalar field, Maxwell's equations for the gauge field and Einstein's equations governing the metric field. We then discuss how the initial states are obtained using the shooting method. Integral and differential expressions in isotropic coordinates are then derived for the conserved charge $Q$ and mass $M$. During the evolution these quantities should remain constant and this allows one to monitor the accuracy of the code at each time step. Expressions for the gravitational and matter Lagrangian as well as the interior Gibbs free energy are then derived. We finally present the thermodynamic results from the numerical simulation.

We work in geometric units $G\eq c\eq 1$ where energy, mass and time have units of length. Coulomb's constant is set to unity so that electric charge has units of length also. The metric has signature $(-1,1,1,1)$. Spacetime indices are in Greek and run from $0$ to $3$. 

\subsection{Reissner-Nordstr\"{o}m metric in isotropic coordinates}

During our numerical simulation in isotropic coordinates, we expect the metric to settle to the Reissner-Nordstr\"{o}m (RN) metric at late times. We  therefore need to express the RN metric in isotropic coordinates.  A general spherically symmetric time-dependent $4D$ metric in isotropic coordinates takes the form \cite{Finelli1,Finelli2,B-E}: 
\beq
ds^2= -N(r,t)^2 dt^2 +\psi(r,t)^4 (dr^2 + r^2 d\Omega^2) 
\eeq{isometric}
where $\psi(r,t)$ is referred to as the conformal factor and $N(r,t)$ is called the lapse function. Note that the isotropic radial coordinate $r$ is not the areal radius. We assume asymptotic flatness so that $N$ and $\psi$ are both unity at infinity. Note that the coordinate time $t$ coincides with the time measured by a clock at rest at infinity. Our goal is to find the analytical expressions for $N$ and $\psi$ that correspond to the RN metric. In standard coordinates, the RN metric is given by \cite{Poisson}
\beq
ds^2= -\Big(1-\dfrac{2M}{r'} + \dfrac{Q^2}{r'^{\,2}}\Big)\,dt^2 + \Big(1-\dfrac{2M}{r'}+\dfrac{Q^2}{r'^{\,2}}\Big)^{-1}\, dr'^{\,2} +r'^{\,2}\,d\Omega^2  
\eeq{RN}
where $M$ and $Q$ are the mass and charge of the black hole and $r'$ is the areal radius. The function $f(r')\eq 1-2M/r' + Q^2/r'^{\,2}$ has zeroes at $r'_{\pm}= M \pm \sqrt{M^2-Q^2}$ where $r'_+$ and $r'_{-}$ are referred to as the outer and inner horizon respectively. Matching the metric \reff{isometric} to the metric \reff{RN} yields $\psi^4= r'^{\,2}/r^2$, $N^2 =f(r')$ and the equation $\tfrac{dr^2}{r^2}= \tfrac{1}{f(r')}\tfrac{dr'^{\,2}}{r'^{\,2}}$. With the condition that $\psi$ and $N$ are unity asymptotically, the latter has solution $r'\eq r+ M + (M^2-Q^2)/4r$ so that the {\it exterior} region of the RN metric in isotropic coordinates is given by 
\beq
ds^2=- \dfrac{\Big(1-\dfrac{M^2-Q^2}{4\,r^2}\Big)^2}{\Big(1 +\dfrac{M}{r} +\dfrac{M^2-Q^2}{4r^2}\Big)^2} \,\,dt^2 + \Big[1 +\dfrac{M}{r} +\dfrac{M^2-Q^2}{4r^2}\Big]^2 \Big( dr^2 + r^2\,d\Omega^2\Big) \,.
\eeq{IsoRN}    
The outer horizon in isotropic coordinates is situated at $r_+\eq \sqrt{M^2-Q^2}/2$. At this location, the lapse function is zero: $N(r_+)\eq0$. The metric \reff{IsoRN} does not cover the interior region of the RN black hole; it covers the static exterior region twice. The minimum value of $r'\eq r+ M + (M^2-Q^2)/4r\,$ is $\,r'_+\eq M+\sqrt{M^2-Q^2}$ ; in metric \reff{IsoRN} both the region $r\ge r_+$ and $r \le r_+$  correspond to the exterior region $r'\ge r'_+$. The Killing vectors in the interior region between the two horizons of the RN black hole are all spacelike \cite{C-E} and the spacetime is nonstationary in this region. To cover the interior region of the RN black hole in isotropic coordinates the functions $N$ and $\psi$ in \reff{isometric} must be time-dependent. In standard coordinates, the spacetime becomes nonstationary when one crosses the outer horizon into the interior because the function $f(r')$ switches sign and the radial coordinate becomes timelike and the time coordinate becomes spacelike. In isotropic coordinates, the metric coefficients in \reff{isometric} are positive-definite; they do not switch sign but instead become time-dependent in the interior. During our numerical simulation, the metric functions $N(r,t)$ and $\psi(r,t)$ in the isotropic metric \reff{isometric} are nonstationary in the region $r\!<\!r_+$, reflecting the true nature of the interior spacetime. It is only in the exterior region $r\!>\!r_+$ that the metric approaches the static form \reff{IsoRN} at late times.

We do not observe the second (inner) horizon or timelike singularity associated with the RN metric \reff{RN}; we observe one horizon at $r_+$ (where $N(r_+)=0$) and a spacelike singularity as in the Schwarzschild case. This is in accord with the findings of previous numerical work on charged collapse \cite{Brady,Hod,Piran}. The inner horizon of the RN metric \reff{RN} is an artifact of exact staticity (and exact spherical symmetry) \cite{Poisson} and it has been known since pioneering work in the 90's \cite{Poisson-Israel}, that the inner (Cauchy) horizon is unstable to perturbations.

\section{Evolution and constraint equations in isotropic coordinates}

\subsection{Matter sector}

For matter, we consider a complex (charged) scalar field coupled to an electromagnetic field $A_{\mu}$. The matter Lagrangian density $\mathcal{L}_{m}$ has a local $U(1)$ gauge symmetry and is given by \cite{Schroeder}   
\begin{equation}\label{L_Matter}
	\mathcal{L}_{m}=-\frac{1}{2}\left(\chi_{;\,\mu}+ieA_\mu\chi\right)g^{\mu\nu}\left(\overline{\chi}_{;\,\nu}-
	ieA_\nu\overline{\chi}\right)-\dfrac{1}{16\pi}F_{\mu\nu}\,F^{\mu\nu}
\end{equation}
where a semi-colon denotes covariant differentiation evaluated with metric \reff{isometric}, a bar denotes complex conjugation and $F_{\mu\nu}\!\equiv \!A_{\nu;\,\mu}-A_{\mu;\,\nu}$ is the electromagnetic field tensor. Spherical symmetry reduces the number of gauge components from four to two: only $A_t=A_{0}$ and $A_{r}=A_{1}$ are non-zero. Gauge freedom allows one to further eliminate $A_r$. This leaves $A_t$ as the only non-zero component and for simplicity we label it $a$. The matter fields are therefore $\chi\eq \chi(r,t)$ and $a\eq a(r,t)$. Lagrange's equations of motion for matter are 
\begin{equation}\label{Poisson}
	\nabla_\alpha\frac{\partial\mathcal{L}_m}{\partial q_{;\,\alpha}}-\frac{\partial\mathcal{L}_m}{\partial q}=0
\end{equation}
where $q$ is a generic field.

\subsubsection{Equations of motion for scalar field $\chi$}

Applying \reff{Poisson} to the scalar field $\chi$ yields the wave equation    
\begin{equation}
	\chi_{;\,\mu\nu}\,g^{\mu\nu}+ieA_\mu\,g^{\mu\nu}\left(2\chi_{;\,\nu}+ieA_{\nu}\chi\right)+ieA_{\mu;\,\nu}\,g^{\mu\nu}\chi=0 \,.
\label{KG}\end{equation}
With spherical symmetry, the three terms in the above equation reduce to
\begin{align*}	
&\chi_{;\,\mu\nu}\,g^{\mu\nu}=\dfrac{1}{\sqrt{-g}}\partial_\mu\left(\sqrt{-g}\,g^{\mu\nu}\partial_\nu\chi\right)
														=-\dfrac{1}{N\psi^6}\partial_t\left(\dfrac{\psi^6\dot{\chi}}{N}\right)
														+\dfrac{1}{N\psi^6r^2}\partial_r\left(N\psi^2r^2\chi'\right)\\
&ieA_\mu \,g^{\mu\nu}\left(2\chi_{;\,\nu}+ieA_{\nu}\chi\right)=-\dfrac{iea}{N^2}\left(2\dot{\chi}+iea\chi\right)\\
&ieA_{\mu;\,\nu}\,g^{\mu\nu}\chi = -\frac{1}{N\psi^6}\partial_t\left(\frac{iea\chi\psi^6}{N}\right)+\frac{iea\dot{\chi}}{N^2}\,
\end{align*}
where $\dot{\chi}\equiv \partial\chi/\partial t$ and $\chi'\equiv \partial \chi/\partial r$. 
Equation \reff{KG} now reads
\beq
-\dfrac{1}{N\psi^6}\,\partial_t\left[\dfrac{\psi^6}{N}\left(\dot{\chi} +iea\chi\right)\right] + \dfrac{1}{N\psi^6r^2}\,\partial_r\left(N\psi^2r^2\chi'\right) -\dfrac{iea}{N^2}(\dot{\chi} +iea\chi)=0\,.
\eeq{KG22}
For numerical purposes, we split the above second order equation into two first order evolution equations. To this end we define the quantity
\begin{equation}\label{p}
	p\equiv\frac{\psi^6}{N}\left(\dot{\chi}+iea\chi\right)\,.
\end{equation}
 In terms of $p$, equation \reff{KG22} simplifies to  
\begin{equation}\label{wave}
	\dot{p}=\frac{1}{r^2}\,\partial_r\left(N\psi^2r^2\chi'\right)-ieap.
\end{equation}
Rearranging \reff{p} we obtain
\begin{equation}\label{e2}
	\dot{\chi}=\frac{N}{\psi^6}\left(p-\frac{iea\psi^6\chi}{N}\right).
\end{equation}
Equations \reff{wave} and \reff{e2} are the evolution equations for $p$ and $\chi$ respectively. 

\subsubsection{Equations of motion for gauge field $a$}

Applying \reff{Poisson} to the gauge field $A_{\mu}$ yields Maxwell's equations: 
\begin{equation}
	\frac{1}{2\pi}F_{\mu\nu;\rho}\,g^{\nu\rho}+ie\chi\left(\overline{\chi}_{;\mu}-ieA_\mu\overline{\chi}\right)
	-ie\overline{\chi}\left(\chi_{;\nu}+ieA_\mu\chi\right)=0.
\end{equation}
Spherical symmetry reduces the above to two equations, one for $\mu=t$ and $\mu=r$. The equation for $\mu=t$ is  
\begin{equation}
		-\frac{a''}{\psi^4}+\frac{N'a'}{N\psi^4}-\frac{2\psi'a'}{\psi^5}-\frac{2a'}{\psi^4r}
		+\frac{2\pi ieN}{\psi^6}\left(\chi\,\overline{p}-\overline{\chi}\,p\right)=0
\end{equation}
where $p$ was defined in \reff{p}. The above can be rewritten as
\begin{equation}\label{M1}
		\frac{1}{r^2}\partial_r\left(\frac{\psi^2a'r^2}{N}\right)=2\pi ie\,\left(\chi\,\overline{p}-\overline{\chi}\,p\right)\,.
\end{equation} 
The above is a constraint equation for the gauge field $a$ (one can view it as Poisson's equation).
Maxwell's equation for $\mu=r$ yields
\begin{equation}
		-\frac{\dot{a}'}{N^2}+\frac{\dot{N}a'}{N^3}-\frac{2\dot{\psi}a'}{N^2\psi}
		+2\pi ie\,\chi\,\overline{\chi}'-2\pi ie\,\overline{\chi}\,\chi'=0
\end{equation}
which can be expressed as
\begin{equation}\label{M2}
	\dot{g}=2\pi ieN\psi^2\left(\chi\,\overline{\chi}'-\overline{\chi}\,\chi'\right)
\end{equation}
with \begin{equation}\label{g}
	g\equiv\frac{a'\psi^2}{N}\,.
\end{equation}
Equation \reff{M2} is the evolution equation for $g$. The equation governing the gauge field $a$ is not an evolution equation but an ordinary differential equation: $a'= g\,N/ \psi^2$.

\subsection{Gravitational sector} 

\subsubsection{Stress-energy tensor}
The stress-energy tensor $T_{\mu\nu}$ appearing in Einstein's field equations can be calculated from the matter Lagrangian \reff{L_Matter}. We first express the latter in terms of the new quantities $p$ and $g$ defined in \reff{p} and \reff{g} respectively:
\begin{equation}
\mathcal{L}_m=\frac{p\overline{p}}{2\psi^{12}}-\dfrac{\chi'\,\overline{\chi}'}{2\psi^4}+\dfrac{g^2}{8\pi\psi^8}.
\label{L_Matter2}\end{equation}
 
The energy-momentum tensor is given by
\begin{equation}
	T_{\alpha\beta}=-2\dfrac{\partial\mathcal{L}_m}{\partial g^{\alpha\beta}}+\mathcal{L}_m\,g_{\alpha\beta}
\end{equation}
and its non-zero components are 
\begin{equation}
	\begin{aligned}
		T_{\theta\theta} &=\dfrac{1}{2}\left(\dfrac{p\overline{p}}{\psi^{8}}-\chi'\,\overline{\chi}'
										 +\dfrac{g^2}{4\pi\psi^4}\right)r^2
		&T_{\phi\phi} 	  & =T_{\theta\theta}\sin^2\theta\\
		T_{rr} &=\dfrac{1}{2}\left(\dfrac{p\overline{p}}{\psi^{8}}+\chi'\,\overline{\chi}'
										 -\dfrac{g^2}{4\pi\psi^4}\right)
		&T_{rt}				&	 =\dfrac{N}{2\psi^6}\left(\chi'\overline{p}+\overline{\chi}'p\right)\\
		T_{tt}					 &=\dfrac{N^2}{2}\left(\frac{p\overline{p}}{\psi^{12}}+\dfrac{\chi'\overline{\chi}'}{\psi^4}
										 +\dfrac{g^2}{4\pi\,\psi^8}\right).
	\end{aligned}
\end{equation}

\subsubsection{Field equations}
Einstein's field equations are given by $G_{\mu\nu}=\kappa^2\,T_{\mu\nu}$
where $\kappa^2\equiv8\pi\,G \eq 8\pi$ and $G_{\mu\nu}$ is the Einstein tensor evaluated with metric \reff{isometric}.   
The $G_{rr}$ equation yields evolution equations for the conformal factor $\psi$ while $G_{tt}$ and $G_{rt}$ yield constraint equations.  $G_{\theta\theta}$ yields an ordinary differential equation for the lapse function $N$ ($G_{\phi\phi}$ yields the same equation).    

The $G_{rr}$ equation is
\begin{equation}\label{Grr}
	\begin{aligned}	&\frac{2}{r\psi^2N^3}\left(2r\psi'^2N^3-4r\dot{\psi}^2N\psi^4+2\psi'N^3\psi+2r\dot{N}\dot{\psi}\psi^5-2r\ddot{\psi}N\psi^5
  +2rN'\psi'N^2\psi\right.\\
  &\left.+N'N^2\psi^2\right)=\dfrac{\kappa^2}{2}\left(\frac{p\overline{p}}{\psi^{8}}
  +\chi'\overline{\chi}'-\dfrac{g^2}{4\pi \psi^4}\right).
  \end{aligned}
\end{equation}
We split the above second order equation into two first order evolution equations. For this purpose we define 
\begin{equation}\label{K}
	K\equiv  -\dfrac{6 \dot{\psi}}{N\psi}\,.
\end{equation}
$K$ is in fact the negative of the trace of the extrinsic curvature for spacelike hypersurfaces at constant time $t$. Equation \reff{Grr} can now be expressed as   
\begin{equation}\label{kev}
	\begin{aligned}
		\frac{\dot{K}}{N}&=\dfrac{K^2}{2}-\dfrac{6\psi'}{\psi^5}\left(\dfrac{\psi'}{\psi}+\frac{1}{r}\right)
											-\dfrac{3N'}{N\psi^4}\left(\dfrac{2\psi'}{\psi}+\dfrac{1}{r}\right)\\
											&+\dfrac{3\kappa^2}{4}\left(\dfrac{\chi'\overline{\chi}'}{\psi^4}+\dfrac{p\overline{p}}{\psi^{12}}
											-\dfrac{g^2}{4\pi \psi^8}\right)\,.
	\end{aligned}
\end{equation}
Equations \reff{K} and \reff{kev} are evolution equations for $\psi$ and $K$ respectively.  
The $G_{tt}$ equation is 
\begin{equation*}
	\dfrac{4}{r\psi^5}\left(3\dot{\psi}^2\psi^3r-2\psi'N^2-N^2r\psi''\right)=
	\dfrac{\kappa^2N^2}{2}\left(\frac{p\overline{p}}{\psi^{12}}+\frac{\chi'\overline{\chi}'}{\psi^4}
	+\dfrac{g^2}{4\pi \psi^8}\right)
\end{equation*}
and using $K$ can be cast in the form 
\begin{equation}\label{c1}
	-\dfrac{4}{\psi^5r}\left(2\psi'+r\psi''\right)
											=\dfrac{\kappa^2}{2}\left(\dfrac{\chi'\overline{\chi}'}{\psi^4}+\dfrac{p\overline{p}}{\psi^{12}}
											+\dfrac{g^2}{4\pi\psi^8}\right)
											-\dfrac{K^2}{3}.
\end{equation}
Equation \reff{c1} is the energy constraint. 
The $G_{rt}$ equation is
\begin{equation*}	
	\dfrac{4}{\psi^2N}\left(\dot{\psi}\psi'N-\dot{\psi}'N\psi+\dot{\psi}N'\psi\right)
	=\kappa^2\frac{N}{2\psi^6}\left(\chi'\overline{p}+\overline{\chi}'p\right)\,
\end{equation*}
which can be expressed as 
\begin{equation}\label{Grt}
	\frac{K'}{3}=\frac{\kappa^2}{4\psi^6}\left(\chi'\overline{p}+\overline{\chi}'p\right)\,.
\end{equation}
Equation \reff{Grt} is the momentum constraint.
The $G_{\theta\theta}$ equation gives
\begin{equation*}
	\begin{aligned}
	&\dfrac{r}{\psi^2N^3}\left(4r\psi^5\dot{\psi}\dot{N}-8r\psi^4\dot{\psi}^2N-4r\psi^5\ddot{\psi}N+N'\psi^2N^2
	-2r\psi'^2N^3+2r\psi'N^3\psi\right.\\ 
	&\left.+2rN^3\psi\psi''+r\psi^2N^2N''\right)
	=\dfrac{\kappa^2}{2}\left(\dfrac{p\overline{p}}{\psi^{8}}-\chi'\overline{\chi}'
	+\dfrac{g^2}{4\pi \psi^4}\right)r^2\,.
	\end{aligned}
\end{equation*}
Eliminating the time derivatives using the $G_{rr}$ equation \reff{Grr}, the above equation reduces to an ordinary differential equation for $N$ 
\begin{equation}\label{Gthetatheta}
	\dfrac{2r}{\psi^2}\partial_r\left(\frac{\psi'}{r\psi^3}\right)+\dfrac{r}{N}\partial_r\left(\frac{N'}{r\psi^4}\right)
	=-\dfrac{\kappa^2\chi'\overline{\chi}'}{\psi^4}+\dfrac{\kappa^2g^2}{4\pi\psi^8}\,
\end{equation}

The evolution equation \reff{kev} for $K$ can be made more numerically stable by combining it with \reff{c1} to obtain
\begin{equation}
	\begin{aligned}
		\frac{\dot{K}}{N}&=K^2-\frac{6\psi'}{\psi^5}\left(\dfrac{\psi'}{\psi}+\frac{1}{r}\right)
		-\dfrac{3N'}{N\psi^4}\left(\dfrac{2\psi'}{\psi}+\dfrac{1}{r}\right)\\
		&-\dfrac{6}{r\psi^5}\left(2\psi'+r\psi''\right)-\dfrac{3\kappa^2g^2}{8\pi\psi^8}.
	\end{aligned}
\label{kev2} \end{equation}

To summarize, the evolution equations for the matter sector are \reff{e2} and \reff{p}  for the pair ($\chi$,$p$) and \reff{M2} for $g$ respectively. The evolution equations for the gravitational sector are \reff{K} and \reff{kev2} for the pair ($\psi,K$). The gauge field $a$ and lapse function $N$ obey the ordinary differential equations \reff{g} and \reff{Gthetatheta} respectively. There is one constraint equation in the matter sector, namely ``Poisson's" equation \reff{M1}. In the gravitational sector, the energy and momentum constraint are given by equations \reff{c1} and \reff{Grt} respectively. Once the initial states and boundary conditions are specified, the evolution of the fields is unique.

\section{Initial states}

Let $\chi_1$ and $\chi_2$ be the real and imaginary part of the complex scalar field $\chi$. We begin by choosing the initial configuration for $\chi_1$, $\chi_2$, $\dot{\chi_1}$ and $\dot{\chi_2}$. It is convenient to pick a configuration such that the momentum constraint \reff{Grt} is trivially satisfied i.e. where the right hand side of \reff{Grt} is zero so that $K'=0$ initially. Since $K=0$ asymptotically, this implies that $K$ is initially zero everywhere. We use three different profiles that satisfy this property: 
\begin{center}
	\begin{tabular}{ p{1 cm}   p{6 cm}  p{4 cm} }
			I:				
			&$\chi_1=\chi_2=\frac{8\lambda_1r^4}{\left(\lambda_1^2+r^2\right)^4}$
			&$\dot{\chi_1}=-\dot{\chi_2}=\frac{\lambda_2\chi_1}{8\lambda_1}$\\[0.3 cm]
			II:
			&$\chi_1=\chi_2=\lambda_2\left(\text{tanh}\left(\lambda_1-r\right)+1\right)$
			&$\dot{\chi_1}=-\dot{\chi_2}=\frac{\lambda_3\chi_1}{\lambda_2}$\\[0.3 cm]
			III:
			&$\chi_1=\frac{8\lambda_1r^4}{\left(\lambda_1^2+r^2\right)^4}\:\:\:\:;\:\:\:\:\chi_2=0$
			&$\dot{\chi_1}=\dot{\chi_2}=0.$\\		
	\end{tabular}
\end{center}
Profile $III$ represents a real scalar field with no charge.
One is free to choose the values of the different $\lambda_i$ and these in turn determine the values of the conserved mass $M$ and charge $Q$. The initial states for the metric functions $\psi$ and $N$ and for the gauge field $a$ are obtained by solving three coupled second order differential equations: the energy constraint \eqref{c1}, the ordinary differential equation \eqref{Gthetatheta} and ``Poisson's" equation \eqref{M1}. These can be split into six first order differential equations:  
\begin{equation}\label{SixEq}
	\begin{aligned}
		&A'=-\frac{\psi^5r^2\kappa^2}{8}\left(\frac{\chi'\overline{\chi}'}{\psi^4}
		+\frac{\left(\dot{\chi}+iea\chi\right)\left(\dot{\overline{\chi}}-iea\overline{\chi}\right)}{N^2}
		+\frac{C^2}{4\pi r^4\psi^8}\right)\\
		&B'=\frac{N}{r}\left(\frac{6A}{r^3\psi^5}\left(1+\frac{A}{r\psi}\right)
		+\frac{\kappa^2\left(\dot{\chi}+iea\chi\right)\left(\dot{\overline{\chi}}-iea\overline{\chi}\right)}{4N^2}
		-\frac{\kappa^2}{4\psi^4}\left(3\chi'\overline{\chi}'-\frac{5C^2}{4\pi r^4\psi^4}\right)\right)\\
		&C'=\frac{2\pi ir^2\psi^6}{N}\left(\chi\dot{\overline{\chi}}-\overline{\chi}\dot{\chi}-2iea\chi\overline{\chi}\right)\\
		&\psi'=\frac{A}{r^2}\\
		&N'=r\psi^4B\\
		&a'=\frac{CN}{r^2\psi^2}
	\end{aligned}
\end{equation}
where the last three equations introduce the new variables $A,B$ and $C$: 
\begin{equation}
A\equiv r^2\psi' \quad;  \quad B\equiv \frac{N'}{r\psi^4}\quad ; \quad C\equiv\frac{r^2\psi^2a'}{N}\,.
\end{equation}
The functions range from 0 to a large distance $R$, the outer computational boundary representing ``infinity". The spacetime is asymptotically flat and boundary conditions at $r\eq R$ that are consistent with this are $N(R)\eq 1$, $\psi(R)\eq 1$ and $B(R)\eq 0$. Gauge invariance allows us to set $a(R)=0$. The initial configuration is non-singular at $r\eq 0$ and this implies that $A(0)\eq 0$ and $C(0)\eq 0$. To solve equations \reff{SixEq} with the above boundary conditions, we use a shooting method with a point located between the origin and $R$. From the boundary conditions, two fields are known at the origin and four fields are known at $R$. We make an educated guess as to the value of each field at the missing end (a total of six numbers.) We then evolve the fields from both ends toward the middle point using fourth order Runge-Kutta. The goal is to modify the six unknown numbers until the fields match in the middle. This can be achieved in a rather straightforward way using a six dimensional Newton method \cite{Press}. The system of linear algebraic equations can be solved using an lower/upper matrix decomposition \cite{Press}. A first estimate for the fields can be obtained by solving the equations for $\dot{\chi}=0$ which applies to a  scalar field. The fields can be made to fit at the middle point to within a few parts in $10^{15}$. We then increase $\dot{\chi}$ by a small amount and use the preceding values of the fields to estimate the new ones. We repeat the procedure until a high enough charge to mass ratio has been obtained. It is worthwhile to note that if the fields match at the middle point, it can be shown that their derivatives also match.

\section{Expressions for the conserved charge $Q_{tot}$ and mass $M_{tot}$\\ in isotropic coordinates}

In this section we obtain expressions in isotropic coordinates for the total (conserved) charge $Q_{tot}$ and mass $M_{tot}$. During the simulation, these quantities should remain constant and this is used to monitor the simulation. The mass $M$ and charge $Q$ of the black hole itself is significantly less (magnitude wise) than $M_{tot}$ and $Q_{tot}$ respectively because a considerable amount of charge and mass is expelled during the collapse process. $M$ and $Q$ will be evaluated in the results section with a different expression. For the analytical stationary RN metric \reff{RN} or \reff{IsoRN}, there is no outgoing matter wave so that $M$ and $Q$ are equal to $M_{tot}$ and $Q_{tot}$ respectively. We leave them expressed in terms of $M$ and $Q$ because in the results section they are evaluated with our black hole values of $M$ and $Q$ (not with our values of $M_{tot}$ or $Q_{tot}$ as these do not correspond to the mass and charge of the black hole respectively).  

The total conserved charge $Q_{tot}$ is given by the general formula \cite{Poisson}
\beq
	Q_{tot}=\dfrac{1}{8\pi}\oint_S F^{\alpha\beta}\,dS_{\alpha\beta}
\eeq{EQ1}
where $S$ is a closed two-surface bounding the charge distribution and $dS_{\alpha\beta}$ is the two dimensional surface element given by \cite{Poisson}
\begin{equation}
	dS_{\alpha\beta}=-2n_{[\alpha}r_{\beta]}\sqrt{\sigma}d^2\theta \,.
\end{equation}
$n^{\alpha}$ and $r^{\alpha}$ are the timelike and spacelike normals to the two-surface and the square brackets denote antisymmetrization: $n_{[\alpha}r_{\beta]}= (n_{\alpha}r_{\beta}-n_{\beta}r_{\alpha})/2$. The induced surface element on $S$ is $\sqrt{\sigma} \,d^2\theta$ where $\sigma$= det($\sigma_{AB}$) where $\sigma_{AB}$ is the induced metric on $S$ (A,B is either $\theta$ or $\phi$). For the metric \reff{isometric}, we obtain $n_\alpha=\left(-N,0,0,0\right)$ and $r_\beta=\left(0,\psi^2,0,0\right)$. The only non-zero components of $F^{\alpha\beta}$ are 
$F^{tr}=-F^{rt}= \tfrac{a'}{N^2\,\psi^4}$. The only non-zero components of $dS_{\alpha\beta}$ are $dS_{tr}\eq -dS_{rt}\eq N\,\psi^6 r^2 \sin\theta d\theta d\phi$. Putting these results together we obtain an expression for the total charge:
\begin{equation}
	Q_{tot}=\dfrac{\psi^2r^2a'}{N}\big|_{r=R}=a'\,R^2\big|_{r=R}
\label{QQ}\end{equation}
where we used the fact that asymptotically, in the large $R$ (infinite) limit, $N$ and $\psi$ approach unity. Using the constraint equation \reff{M1}, one can express the charge $Q_{tot}$ in the integral form
\beq
Q_{tot}= \int_0^R  2\pi ie\,\left(\chi\,\overline{p}-\overline{\chi}\,p\right) \,r^2 dr \,.
\eeq{QInt}
From the above expression, one can define an effective charge density  
\beq
\rho= \dfrac{ie}{2}\,\left(\chi\,\overline{p}-\overline{\chi}\,p\right)\,.
\eeq{rho2}
  
The total (conserved) ADM mass $M_{tot}$ for an asymptotically flat spacetime is given by \cite{Poisson}
\beq
	M_{tot} = -\frac{1}{8\pi} \oint_{S} (k-k_0) \sqrt{\sigma}\,d^2\theta
\eeq{ADMmass}
where $S$ is the two-sphere at spatial infinity. $k$ is the trace of the extrinsic curvature of $S$ embedded in $\Sigma_t$, the three-dimensional spacelike hypersurface at constant $t$.  $k_0$ is the trace of the extrinsic curvature of $S$ embedded in flat spacetime. $k$ and $k_0$ can be readily evaluated and one obtains $k-k_0 =4\psi'/\psi^3$. The mass $M_{tot}$ is equal to
	\beq
		M_{tot} = -2 r^2 \psi' \,\psi \big|_{r=R} = -2 R^2 \,\psi' \big|_{r=R} 
		\eeq{ADMMass}
where we used that $\psi=1$ asymptotically.      
Using \reff{c1} one can express the mass in the integral form 
\begin{equation}
	M_{tot}	=\dfrac{\kappa^2}{4}\int_0^R\left(\dfrac{\chi'\overline{\chi}'}{\psi^4}+\dfrac{p\overline{p}}{\psi^{12}}
		+\dfrac{g^2}{4\pi\psi^8}-\dfrac{2K^2}{3\kappa^2}\right)r^2\psi^5dr.
\end{equation}

\section{Free energy}
There are arguments, based on the Euclidean action \cite{Hawking2}, for relating the negative of the total Lagrangian of a stationary black hole to its free energy. As far as we know, to date, there is no direct analytical proof of this; it has only been tested numerically during gravitational collapse to a Schwarzschild black hole \cite{Khlebnikov, C-E}. One of the goals of this paper is to test this association numerically for the case of charged collapse where the relevant thermodynamic potential is the Gibbs free energy \cite{Hawking2, Kiritsis}. To do so, we first need to determine an expression for the Lagrangian.

\subsection{Gravitational and Matter Lagrangian}

The gravitational action $S_G[g]$ is a sum of the Einstein-Hilbert term $S_H[g]$, a boundary term $S_B[g]$ and a nondynamical term $S_0$ \cite{Poisson}:
	\beq
		S_{G}[g] = S_H[g]+ S_B[g] + S_0 = \int L_G \,dt 
	\eeq{SG}
where $L_G$ is the gravitational Lagrangian and 
\beq
S_H[g]= \frac{1}{16\pi} \int \sqrt{-g}\,R \, d^3x\, dt \,.\\
\eeq{SH}
The Ricci scalar in isotropic coordinates is given by    
	\beqa
		R &=& \frac{2}{rN^3\psi^5} \big( 18r\dot{\psi}^2\,N\psi^3 + 6r\ddot{\psi}N\psi^4 - 8\psi'N^3 - 4r\psi''N^3 \\
		& & {} - 6r\dot{N}\dot{\psi}\psi^4 - rN''N^2\psi - 2rN'\psi'N^2 - 2N'N^2\psi \big)\nonumber\,.
	\eea
The Ricci scalar is not the Lagrangian density as it contains second derivatives of the metric tensor. Besides the Hilbert action $S_H[g]$, one also requires the boundary term $S_B[g]$ to obtain Einstein's field equations \cite{Poisson}. $S_0$ ensures that the action $S_G$ is finite and equal to zero in flat spacetime. We do not need to spell out $S_B$ (or $S_0$) to determine the gravitational Lagrangian. As in \cite{Khlebnikov}, $L_G$ can be  obtained simply by integrating out by parts the second derivative terms appearing in $R$ and this yields
	\beq
		L_{G} = \frac{1}{4} \int_0^R \Big( 8r^2\psi'N'\psi + 8r^2N\psi'^2 - \frac{24r^2\dot{\psi}^2\psi^4}{N} \Big)\, dr.
	\eeq{LLG}
Note that $L_G$ contains only first derivatives of the metric functions. The matter Lagrangian is given by the integral of \reff{L_Matter2}:
\beq
L_m =\int \mathcal{L}_m \sqrt{-g}\,d^3x = 4\,\pi \int_0^R  \left(\frac{p\overline{p}}{2\psi^{12}}-\dfrac{\chi'\,\overline{\chi}'}{2\psi^4}+\dfrac{g^2}{8\pi\psi^8}\right) \psi^6 \,|\!N\!| \,r^2 \,dr\,.
\eeq{Lm22}          
The total Lagrangian is the sum $L_m +L_G$. 

\subsection{Gibbs free energy of the Reissnner-Nordstr\"{o}m black hole}  

We can easily calculate the Gibbs free energy $G \eq E\m TS\m \Phi_H Q$ of the RN black hole ($E$ is the ADM mass $M$, $T$ and $S$ are the temperature and entropy respectively and $\Phi_H$ the electrostatic potential at the horizon). The temperature is given by $T=\hbar\,\kappa/2\pi$ where $\kappa$ is the surface gravity of the black hole and $S= A/4\hbar$ where $A$ is the area of the event horizon. For a RN black hole, $\kappa= \sqrt{M^2-Q^2}/{r'_+}^2$ and the area of the horizon is $A\eq 4\pi {r'_+}^2$, where $r'_{+}= M + \sqrt{M^2-Q^2}$ is the (areal) radius of the outer horizon. Therefore the product $T\,S$ is equal to $\sqrt{M^2-Q^2}/2$. The electrostatic potential at the horizon is given by $\Phi_H=Q\slash r'_{+}$ \cite{Hawking2, Kiritsis}. The Gibbs free energy of the RN black hole is equal to 
\begin{equation}
\begin{split}
G&=M-\dfrac{1}{2}\sqrt{M^2-Q^2}- \dfrac{Q^2}{M+\sqrt{M^2-Q^2}}\\
&=\dfrac{1}{2}\sqrt{M^2-Q^2}.
\end{split}
\eeq{F_RN}  
For $Q\eq 0$, expression \reff{F_RN} reduces to the (Helmholtz) free energy $M/2$ of a Schwarzschild black hole \cite{Khlebnikov, C-E}. Note that though $\hbar$ appears in $T$ and $S$, it does not appear in the product $TS$ and hence does not appear in the expression \reff{F_RN} for the Gibbs free energy.   

\subsubsection{Analytical formulas for the exterior and interior contribution to the free energy} 

There is an interior and exterior contribution to the gravitational Lagrangian $L_G$, the integral given by \reff{LLG}. The analytical expressions for $\psi$ and $N$ in the exterior static region can be extracted from metric \reff{IsoRN}:  
\beq
N= \dfrac{\Big(1-\dfrac{M^2-Q^2}{4\,r^2}\Big)}{\Big(1 +\dfrac{M}{r} +\dfrac{M^2-Q^2}{4r^2}\Big)}\quad;\quad \psi=\Big[1 +\dfrac{M}{r} +\dfrac{M^2-Q^2}{4r^2}\Big]^{1/2} \,.
\eeq{NPsi}  
Inserting the above functions into \reff{LLG} and integrating from an isotropic radial coordinate of zero to the outer horizon $r_+= \sqrt{M^2-Q^2}/2$ yields the exterior contribution $L_{G_{ext}}$ to the gravitational Lagrangian 
\beq 
L_{G_{ext}}= \frac{1}{4} \int_{r_+}^{\infty} \Big( 8r^2\psi'N'\psi + 8r^2N\psi'^2 \Big)\, dr = -\dfrac{1}{2}(M+\sqrt{M^2-Q^2})\,.
\eeq{Lext}
There is no charge residing in the exterior of the RN black hole; there is an electric field but no scalar field. The contribution to the matter Lagrangian \reff{Lm22} in the exterior stems entirely from the electromagnetic part, 
\beq 
L_{M_{ext}}= \dfrac{1}{2} \int_{r_+}^{\infty}\frac{g^2}{\psi^2}N r^2 dr\,=\dfrac{Q^2}{2(M+\sqrt{M^2-Q^2})}\,.
\eeq{Lext}
The exterior contribution to the Gibbs free energy is the negative of the total exterior Lagrangian,
\beq
G_{ext}= - L_{G_{ext}} - L_{M_{ext}} = \sqrt{M^2-Q^2}\,.
\eeq{Gext}
The interior contribution is therefore equal to
\beq
G_{int}= G- G_{ext}= -\dfrac{1}{2}\,\sqrt{M^2-Q^2}=-TS=-r_+   
\eeq{Fint}
where $G$ is the Gibbs free energy of the RN black hole given by \reff{F_RN}. As in the Schwarzschild black hole \cite{Khlebnikov, C-E}, the interior contribution to the free energy is equal to the negative of the product of temperature and entropy i.e. $G_{int}=-T\,S$. We will see that the negative dip in the interior occurs in a thin shell near the radius of the horizon, as in the Schwarzschild case \cite{B-E}. Our numerical results in the interior will be compared to the analytical expression \reff{Fint}. 

\section{Numerical Results}

\subsection{Code and initial state}

We work in geometrized units where $G\eq c \eq 1$. We also set Coulomb's constant to unity.  In these units radius, mass, energy, time and charge have units of length. The coupling constant $e$ has units of inverse length. Its value is arbitrary (there is no experimental value for it in our model). We set it equal to unity and leave the scale unspecified (one can work in centimetres, metres, etc.). One unit of time is the time it takes light to travel a radial distance of one unit in flat spacetime\footnote{If the scale were specified to be $3\times 10^{5}$km, then one unit of time would correspond to one second.}. For the spatial grid, we introduce a new coordinate $x$ such that $r=2x/(1-x)$. We use a constant grid in $x$, with $x$ ranging between $0$ and $1$. This maintains a larger concentration of points in the interior than the exterior. The spatial and time increment are taken to be $\Delta x \eq \Delta t \eq 5\times 10^{-5}$. The fields that are governed by an evolution equation are evolved via a fourth order Adams-Bashforth-Moulton (ABM) scheme \cite{Press}. The fields $N$ and $a$ are governed by an ordinary differential equation and are obtained by iterating backwards starting from infinity using equations \reff{Gthetatheta} and \reff{M1} and their boundary conditions respectively. This is performed at each time step using again ABM. The simulation was carried out for the three different initial profiles discussed in the section on the initial states. The results for profiles I, II and III are presented in tables 1, 2 and 3 respectively. $M$ and $Q$ represent the mass and charge of the black hole itself, whereas $M_{tot}$ and $Q_{tot}$ represent the total (conserved) mass and charge. We will see shortly how $M$ and $Q$ are evaluated. $-L_{int}$ and $G_{int}$ represent the interior Lagrangian and Gibbs free energy at late times respectively. The discussion that follows will focus on one initial state with profile $I\!I$. This will allow us to highlight the essential features of the thermodynamics that are common to all cases. The initial state for profile $I\!I$ amounts to specifying the values of three constants $\lambda_1$, $\lambda_2$ and $\lambda_3$. We will discuss the middle case in table 2: $\lambda_1 \eq 2$, $\lambda_2 \eq 0.11$ and $\lambda_3 \eq 0.032$. The total (conserved) ADM mass and charge corresponding to this initial state is $M_{total}\eq 1.241$ and $Q_{total}\eq -0.842$ respectively\footnote{Switching the sign of $\chi_1$ and $\chi_2$ in the initial state leads to a positive charge. The sign has no bearing on the thermodynamics.}. Note that this differs considerably from the mass $M=0.721$ and charge $Q=-0.215$ of the black hole itself because a significant amount of mass and charge are expelled in the collapse process. The total mass and charge are useful for monitoring the accuracy of the code and are checked at each time step. The ADM mass begins to deviate from its original value before the total charge. For the above initial state, the numerical results and plots are obtained up to an evolution time of $t\eq 23.5$, just below $5\%$ ADM mass deviation. The ADM mass is known to be very sensitive to tiny deviations in the metric and matter functions \cite{C-E} i.e. the functions can be evolving well even when the ADM mass begins to deviate from its original value. For example, the numerical curve for the metric function $\psi$ continues to approach the analytical curve of the RN exterior metric even at $5\%$ deviation from the ADM mass.    

\subsection{Plots of metric and matter functions, mass and charge of black hole} 
We made plots of all the relevant functions. The evolution of the conformal factor $\psi$ and lapse $N$ are shown in figures~\ref{psi} and~\ref{n} respectively. We denote $r_+(t)$ as the isotropic radius where the lapse $N$ crosses zero at a time $t$.  Black hole formation is taken to coincide with $N$ crossing zero \cite{Finelli1, Finelli2, Khlebnikov} since radially outgoing (or ingoing) null geodesics in the interior ($r<r_+(t)$) do not cross the $N\eq 0$ spacelike two-surface as it evolves (see figure~\ref{geodesics}). At very late times the $N=0$ spacelike two-surface is identified as the apparent horizon of the stationary RN black-hole (for more details see \cite{B-E}). The radius $r_+(t)$ increases with time and has a numerical value of $0.344$ at $t=23.5$; this is the radius of the apparent horizon. In the interior at late times, the lapse $N$ shows almost no dependence on $r$ and approaches zero asymptotically from below (i.e. $\dot{N} \to 0$ as $N\to 0$). The conformal factor $\psi$ has its peak value at $r_{+}$. In the interior, it also has no spatial dependence and approaches zero asymptotically from above i.e.  $\dot{\psi} \to 0$ as $\psi\to 0$ (see figure~\ref{psidot}). As in \cite{Brady, Hod, Piran} we do not observe an inner horizon; the spacetime structure in the interior resembles that of the uncharged case \cite{C-E}. The absolute value of the complex scalar field $\chi$ is shown in figure~\ref{chi}. It approaches a constant value inside $r_{+}$ and the only region where the time or radial derivatives are high is near the radius $r_{+}$. Note the outgoing matter wave. 

In the RN solution, the electromagnetic-field tensor is given by $F^{tr'}=Q\slash r'^2$ \cite{Poisson} where $r'$ is the areal radius. A short calculation yields $g(r)=Q\slash r^2$ where $g$ is given by \reff{g} and $r$ is the isotropic radius. This, in combination with the relation ${r_+}^2=(M^2-Q^2)\slash 4$, allows to determine the charge and mass of the black hole: 
\beq
	\begin{aligned}
		&Q=r_{+}^2\,g(r_{+})\\
		&M=\sqrt{4{r_{+}}^2+Q^2}\,.
	\end{aligned}
\eeq{MQ}
The above are evaluated at the latest time. 

\subsection{Electrostatic potential and charged shell}
The field $a$ is shown in figure~\ref{a}. Inside $r_{+}$, it is almost perfectly constant in space and its value decreases with time. Between $r_{+}$ and the position of the outgoing wave, it decreases smoothly as $r$ increases. On the other side of the outgoing wave, it decreases faster and this is consistent with the fact that the outgoing wave contains charge. This is what would be expected from basic classical electromagnetism for a situation where two charged shells are concentric and the outside one is moving outward. It can be seen from figure~\ref{rho} that the charge density of the black hole becomes increasingly concentrated near $r_{+}$ as time progresses. At late times it is concentrated just inside the apparent horizon (located at isotropic radius $r=0.344$). In short, at late times one has a charged shell \cite{B-E} with a constant potential $a$ inside. The value of $-a$ extracted at the apparent horizon in our numerical simulation at late times is $-0.174$. The value of $\Phi_H=Q/r'_+$ is $-0.153$ where $r'_+ = M+\sqrt{M^2-Q^2}$ is the areal radius at the horizon. The percentage difference between $-a$ and $\Phi_H$ is around $12\%$, which is explained by the fact that the outgoing wave hasn't reached infinity yet. Note that $\Phi_H$ should be compared to $-a$ since $\Phi_H\equiv-A_{\mu}\,\xi^{\mu}|_H=-A_t|_H$ \cite{Poisson} where $|_H$ means evaluated at the horizon. Here $\xi^{\mu}$ is the Killing vector of the stationary RN black hole and in our coordinate system it is given by $\xi^{\mu}=(1,0,0,0)$. Recall that $a=A_t$ in our numerical simulation. What is important here is that the charge accumulates in a shell at the radial location of the horizon \cite{B-E} and has an electrostatic potential which is numerically close to $\Phi_H$. This provides us with a clear physical interpretation of the work term $\delta W=\Phi_H \delta Q$ in the first law of black hole thermodynamics: it is the work needed to bring a charge $\delta Q$ to the charged shell located at the horizon. Note that though the charged shell is located at the radius of the horizon, the proper radial distance between the shell and the origin decreases towards zero with time because the conformal factor $\psi$ approaches zero in the interior. The charged shell is effectively ``collapsing towards the center" (see \cite{B-E} for details).      

\subsection{Lagrangian and free energy}
\subsubsection{The interior region}
Our goal here is to compare the negative of the total interior Lagrangian $-L$ (gravity + matter) to the interior Gibbs free energy $G_{int}$. The first thing to note is that the matter Lagrangian, plotted in figure~\ref{lm}, makes basically zero contribution to the interior at late times. The interior Lagrangian is entirely gravitational. The integral of the negative of the gravitational Lagrangian from the origin to a given $r$ is shown in figure~\ref{lg}. There is a clear negative dip just inside the apparent horizon that becomes thinner and closer to the horizon with time. Since $G_{int}=-TS$ this suggests that the entropy of the charged black hole accumulates at the horizon and that it is gravitational in origin. In the interior, the metric is not static near the horizon (see figure~\ref{psidot}) but dynamical so that the entropy is associated with the dynamical interior in accord with some previous analytical studies \cite{Vaz1,Vaz2, E-C}. This was also observed numerically for the Schwarzschild case \cite{Khlebnikov,C-E}. The numerical value of the minimum in figure~\ref{lg} is plotted in figure~\ref{lgmin} as a function of time together with the interior Gibbs free energy $G_{int}=-r_+$. Note how the two graphs approach each other with time. At the late time $t=23.5$, the gravitational minimum (which we label $-L_{int}$) is $-0.306$ and $G_{int}=-0.344$, for a difference of $10.9\%$ (see table 2). It is clear from the graphs that the two values would continue to approach each other if the code could evolve further in time. The ADM mass has deviated by $5\%$ at this point. This is due to sharp changes in the gradients in the near-horizon region. 

We verified that the grid size determines how far the simulation runs before the ADM mass begins to deviate from its original value. This is shown in table 4.  We used the following sequence for the number of points: $\{1250,2500,5000,10000,20000\}$ with the corresponding step sizes shown in table 4. The times at which the ADM mass deviated by $1\%$ and the times by which it deviated by $5\%$ are listed. The table shows clearly that decreasing the step size increases the time evolution. The numerical results and plots presented above are for the highest resolution $\Delta x =5.0 \times 10^{-5}$ and are taken at $t=23.5$ just below $5\%$ ADM mass deviation. Going beyond this resolution increases significantly the computing time so that it no longer becomes practical.

The results for all three profiles with different initial states are shown in tables 1, 2 and 3. Note that it is possible to have the total charge $|Q_{tot}|$ be greater than the total mass $M_{tot}$ (see table 1). However, in all cases, $|Q|$ is less than the mass $M$ and there are no naked singularities. The discrepancy between $-L_{int}$ and $G_{int}$ is shown for all cases. As one can see, the discrepancy depends slightly on the initial state. However, in all probability, this difference is not due to fundamental physical reasons. It is simply that the initial state affects when the ADM mass starts to deviate from its original value. A higher resolution would decrease the discrepancy for all initial states and it is expected that all should yield the same thermodynamics. The matter Lagrangian tends to zero in the interior in all cases.  

\subsubsection{Extending time evolution in the exterior region: matter Lagrangian and exterior RN metric}
The results discussed so far are the robust numerical results of this paper. The exterior region has a large outgoing matter wave which can clearly be seen in figure~\ref{lm}. This limits the ability to extract unambiguous numerical results in the exterior. Nonetheless, we still want to check two things in the exterior, at least qualitatively. We would like to verify that due to the presence of the electric field in the exterior, the matter Lagrangian in the exterior does not settle to zero in contrast to the Schwarzschild case \cite{Khlebnikov,C-E}. Secondly, we know the analytical form of the RN metric in the exterior. As a consistency check, we would like to verify that numerically the metric in the exterior approaches it. 

Unlike the near-horizon region, the exterior does not contain regions with very sharp changes in gradients. It is therefore possible to extend the evolution time in the exterior by considering only the points outside of a cutoff where one sets some boundary conditions. The cutoff is chosen here to be beyond $r_+$ at $r=0.5$. The outside evolution starts at $t=23.5$ and we can now relax the time step to $\Delta t =10^{-4}$. The functions that need to be specified at the cutoff are $\psi$, $g$ and the real and imaginary part of $\chi$. However, $\psi$ and $g$ had already plateaued to a constant at the cutoff at $t=23.5$; we therefore set them equal to these respective constants on the boundary. However, $\chi$ is still evolving at $t=23.5$. As time increases, the outgoing wave dissipates and we expect the outside to reach the exterior of the RN black hole which contains no charge density and a zero stress-energy tensor for the scalar field (there is a non-zero stress-energy tensor due to the electric field). This implies that $\chi'$ should tend to zero outside. Since $\chi$ is zero at infinity, the value of $\chi$ at the cutoff should tend to zero. All we know is that it should approach zero but not how it approaches zero as a function of time. Fortunately, the metric is mostly affected by the outgoing wave which is not sensitive to the boundary condition at the cutoff. We therefore impose some time-dependent boundary condition for $\chi$ at the cutoff that leads to a smooth evolution. Figure~\ref{lmat2} shows the matter Lagrangian in the exterior at very large times. It is important to read this graph properly. The wavy part consists of the matter wave which is moving outwards and decreasing in amplitude. What is important is the portion before the wave. From $t=40$ to $t=120$ it remains constant; the matter Lagrangian is clearly plateauing towards a non-zero value. This is due to the external electric field. The conformal factor $\psi$ is plotted in figure~\ref{psiE} together with the analytical form of $\psi$ in the exterior RN metric. At $t=120$, the two curves are very close to each other confirming that the exterior is indeed approaching the Reissner-Nordstr\"{o}m solution.

\section{Conclusion}

In this paper, we showed that the interior Gibbs free energy during charged collapse stems entirely from the gravitational sector and accumulates in the dynamical region inside and near the horizon. Numerically, we showed that the interior Lagrangian agrees with the analytical expectation $G_{int}$ for the interior Gibbs free energy to within roughly $10\%$ depending on the initial state. We observe the formation of a charged shell just inside the horizon with a constant electrostatic potential inside that matches $\Phi_H$ to within $12\%$. The work term $\delta W=\Phi_H\delta Q$ in the first law of black hole thermodynamics can be interpreted as the work required to bring a charge $\delta Q$ to the charged shell at the horizon.  

A consistent dynamical picture of black hole thermodynamics is now emerging from numerical studies of gravitational collapse in isotropic coordinates. First, the Helmholtz free energy $F\eq E \m TS$ in the Schwarzschild case \cite{Khlebnikov, C-E} makes an interior contribution of $F_{int}\eq -TS$ and the Gibbs free energy $G\eq E\m TS -\Phi_H Q$ in the charged case makes also an interior contribution $G_{int}\eq -TS$. The interior contribution in both cases is equal to $-TS$ even though we are dealing with different free energies. Secondly, the interior contribution in both cases stems from the gravitational sector. Thirdly, in both cases it accumulates inside near the horizon. Fourthly, the region where it accumulates is dynamical/non-static. This strongly suggests the following: black hole entropy is gravitational entropy and accumulates in the dynamical interior near the horizon. 

It is now important to check if gravitational collapse to a Kerr black hole obeys these features. Such a study would be a major numerical undertaking. The number of equations to evolve would increase but more importantly, a two dimensional spatial grid would now be required, increasing massively computation time. The excision technique used in numerical simulations of collapse of a rotating neutron star to a Kerr black hole \cite{Baiotti, Hawke} would not be useful in our case. The interior region where the sharp changes in gradients occur in our simulations is precisely the region where we need to extract pertinent numerical results. 

Besides the Kerr black hole, it would be interesting to investigate the thermodynamics of the black holes recently obtained during numerical studies of collapse of a k-essence scalar field \cite{Garfinkle,Gabor}. These fields are exotic in the sense that they have a non-canonical kinetic term \cite{Mukhanov1,Mukhanov2}. It is therefore worthwhile to investigate whether such black holes obey the thermodynamic features discussed above. The equations of motion for k-essence scalar fields coupled to gravity follow from a Lagrangian and this implies that a thermodynamic study should be possible. In particular, one would like to determine if the matter Lagrangian tends towards zero in the interior during collapse as with all previous types of matter investigated to date.    

\clearpage
\begin{table}[ht!]
		\begin{center} 	
  	\begin{tabular}{ p{1.0 cm}   p{1.0 cm}  p{1.3 cm}  p{1.3 cm}  p{1.3 cm} p{1.3 cm} p{1.3 cm} p{1.3 cm} p{2.0 cm} }
    	$\lambda_1$ & $\lambda_2$ & $M_{tot}$ & $Q_{tot}$ & $M$   &   $Q$  	&	$-L_{int}$&	$G_{int}$ &Percentage discrepancy\\ \hline
			1.85				& 8						& 1.364		& -1.234			&	0.644	&	-0.170	&	-0.271		&	-0.311		& 12.6 	\\ 
			1.85				& 12					& 1.669		& -1.959			&	0.638	&	-0.237	&	-0.258		&	-0.296		& 12.8 	\\ 			
			1.8					& 12					& 1.897		& -2.191			&	0.754	&	-0.280	&	-0.304		&	-0.350		& 13.1 	\\ 
  	\end{tabular}
  	\caption{Results for profile I}
  	\end{center}
 	\end{table}

	\begin{table}[ht!]
		\begin{center} 	
  	\begin{tabular}{ p{0.7 cm} p{0.7 cm} p{0.7 cm} p{1.2 cm} p{1.2 cm}  p{1.2 cm} p{1.2 cm} p{1.2 cm} p{1.2 cm} p{2.0 cm} }
    	$\lambda_1$ & $\lambda_2$ & $\lambda_3$	& $M_{tot}$ & $Q_{tot}$	& $M$	& $Q$ & $-L_{int}$ & $G_{int}$	& Percentage discrepancy\\ \hline
			2						& 0.11				&	0.016				&	1.105		& -0.403	&	0.706		&	-0.110	&	-0.315	&	-0.349			& 9.7	\\ 
			2						& 0.11				& 0.032				& 1.241		& -0.842	&	0.721		&	-0.215	&	-0.306	&	-0.344			& 10.9 \\ 
			2						& 0.11				& 0.048				& 1.488		& -1.357	&	0.754		&	-0.307	&	-0.295	&	-0.345			& 14.5 \\ 
  	\end{tabular}
  	\caption{Results for profile II}
  	\end{center}
 	\end{table}

 	\begin{table}[ht!]
  	\begin{center} 	
  	\begin{tabular}{ p{1.0 cm}   p{1.3 cm}  p{1.3 cm} p{1.3 cm} p{1.3 cm} p{2.0 cm} }
    	$\lambda_1$ & $M_{tot}$	& $M$  & $-L_{int}$ & $G_{int}$	& Percentage discrepancy 	\\ \hline
			1.50				& 1.102		&	0.334	 & -0.295			 		&	-0.334			    &	11.5													\\
			1.55				& 0.974		&	0.279	 & -0.244					&	-0.279				  &	12.5													\\			
			1.60				& 0.867		&	0.236	 & -0.203					&	-0.236				  &	14.0													\\			
		\end{tabular}
  	\caption{Results for profile III}
  	\end{center}	  	
 	\end{table}

	\begin{table}[ht!]
		\begin{center} 	
  	\begin{tabular}{ p{3.2 cm}   p{2.21 cm}  p{3.3 cm}  p{3.3 cm}  }
    	Number of spatial steps &$\Delta t$ &Time of deviation of 1\% of $M_{tot}$&Time of deviation of 5\% of $M_{tot}$\\ 	
    	\hline
    	1250				& 0.0008  		&	14.3	& 17.9	\\
			2500				& 0.0004			&	16.4	& 19.4	\\
			5000				& 0.0002			& 18.3	& 20.9	\\
			10000				& 0.0001			& 20.1	& 22.3	\\
			20000				& 0.00005			& 21.7	& 23.7	\\
		\end{tabular}
  	\caption{Evolution time as a function of gridsize for profile $II$ with $\lambda_1=2$, $\lambda_2=0.11$ and $\lambda_3=0.032$}
  	\end{center}
 		\end{table}
\clearpage

\begin{figure}[tbp]
		\begin{center}
			\includegraphics[scale=.38, draft=false, trim=1.5cm 1.5cm 2.5cm 2cm, clip=true]{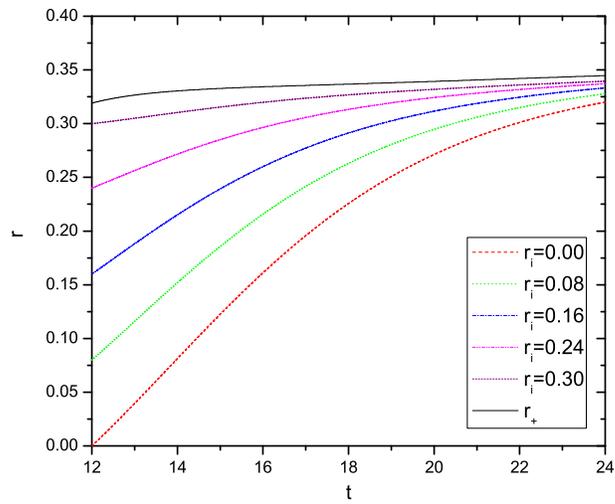}
		\end{center}
		\caption{\label{geodesics} The position of outgoing radial light-like geodesics is shown as a function of time. They are released at $t=12$. The radius $r_+$ where $N=0$ is also shown as a function of time (the top black line). The null geodesics are unable to cross $r_+$ (see also \cite{B-E}).} 
\end{figure}
\clearpage
 
\begin{figure}[tbp]
		\begin{center}
			\includegraphics[scale=.38, draft=false, trim=1.5cm 1.5cm 2.5cm 2cm, clip=true]{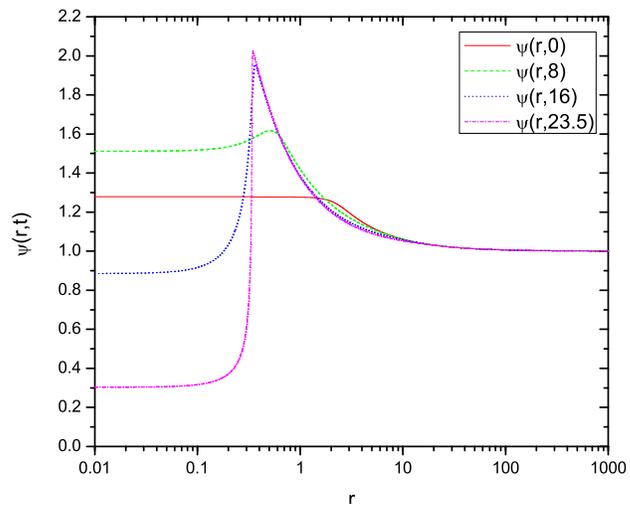}
		\end{center}
		\caption{\label{psi} Space profile of the metric field $\psi$ for different times.}
\end{figure}
\clearpage
\begin{figure}[tbp]
		\begin{center}
			\includegraphics[scale=0.38, draft=false, trim=1.5cm 1.5cm 2.5cm 2cm, clip=true]{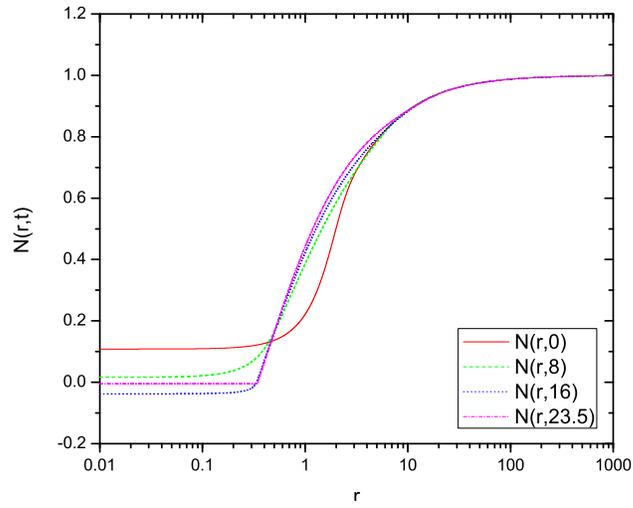}
		\end{center}
	  \caption{\label{n} Space profile of the lapse function $N$ for different times. $r_{+}$ is the radius where $N$ crosses zero at late times. $N$ is negative in the interior and approaches zero from below.}
\end{figure}

\clearpage

\begin{figure}[tbp]
		\begin{center}
			\includegraphics[scale=.37, draft=false, trim=1.5cm 1.5cm 2.5cm 2cm, clip=true]{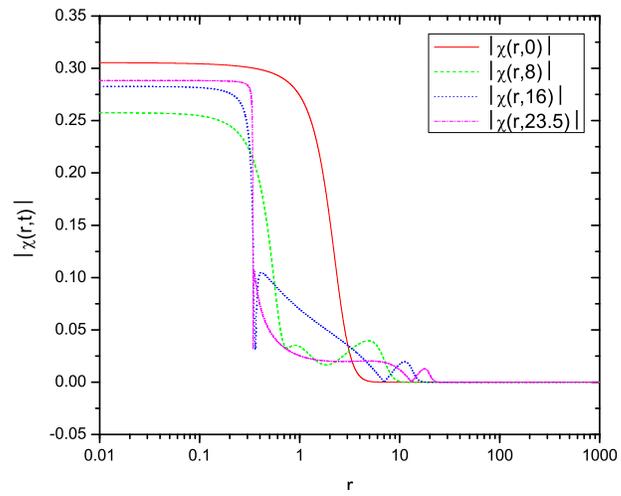}
		\end{center}
		\caption{\label{chi} Space profile of the absolute value of the matter field $\chi$ for different times. Note the propagation of an outgoing matter wave in the exterior region.}
\end{figure}

\clearpage

\begin{figure}[tbp]
		\begin{center}
			\includegraphics[scale=.38, draft=false, trim=1.5cm 1.5cm 2.5cm 2cm, clip=true]{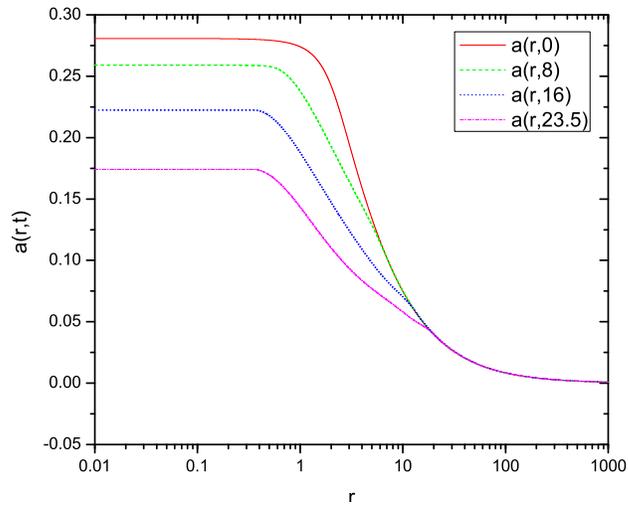}
		\end{center}
		\caption{\label{a} Potential $a$ for different times. Inside the horizon, $a$ is almost constant, consistent with a charged shell configuration at the apparent horizon. The change at large $r$ in the profile of $a$ corresponds to the position of the outgoing wave.}
\end{figure}
	
\clearpage
		
\begin{figure}[tbp]
		\begin{center}
			\includegraphics[scale=.38, draft=false, trim=1.5cm 1.5cm 2.5cm 2cm, clip=true]{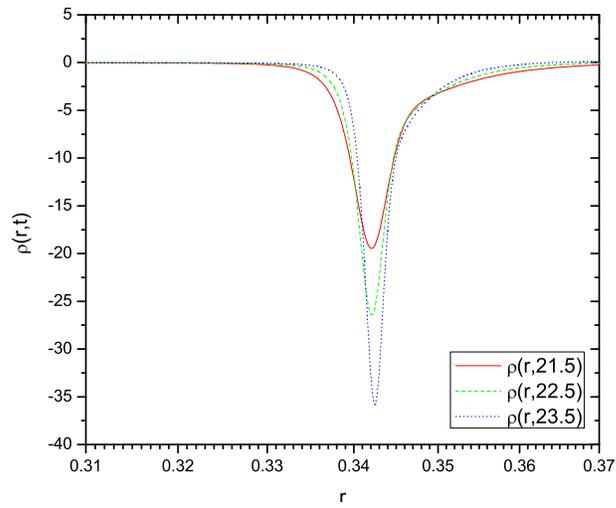}
		\end{center}
		\caption{\label{rho} The charge density at different times. As the system evolves, it becomes more and more concentrated towards the inside of the apparent horizon. At late times one has a charged shell in the vicinity of the horizon.}
\end{figure}		
		
\clearpage
	
\begin{figure}[tbp]
		\begin{center}
			\includegraphics[scale=.38, draft=false, trim=1.5cm 1.5cm 2.5cm 2cm, clip=true]{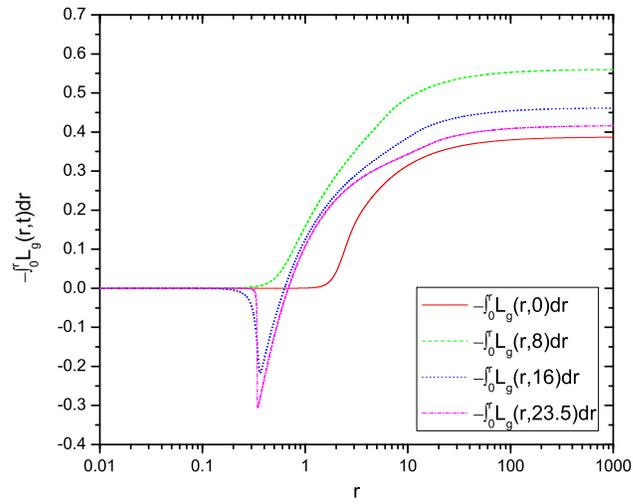}
		\end{center}
		\caption{\label{lg} The gravitational Lagrangian accumulation at different times. Note the negative dip in the thin shell just inside and near the apparent horizon. There is a disturbance in the outgoing wave region.}
\end{figure}

\clearpage
\begin{figure}[tbp]
		\begin{center}
			\includegraphics[scale=.38, draft=false, trim=1.5cm 1.5cm 2.5cm 2cm, clip=true]{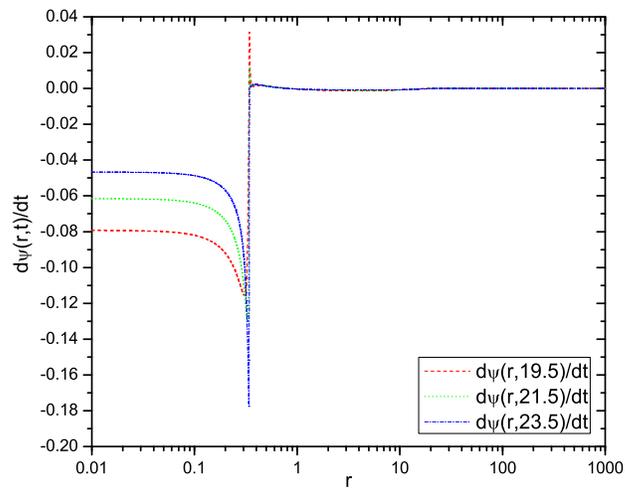}
		\end{center}
		\caption{\label{psidot} The time-derivative of the metric function $\psi$. $\psi$ is static in the exterior and its time-derivative is decreasing towards zero in the interior volume. However, it is not static inside and near the horizon, precisely the region where the interior free energy accumulates.}
\end{figure}

\clearpage
\begin{figure}[tbp]
		\begin{center}
			\includegraphics[scale=.38, draft=false, trim=1.5cm 1.5cm 2.5cm 2cm, clip=true]{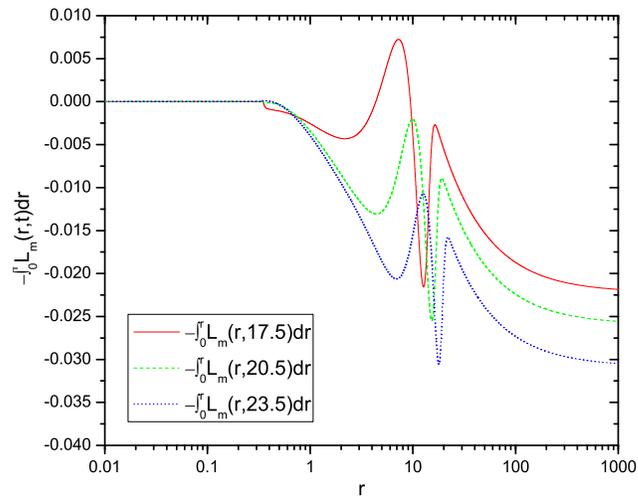}
		\end{center}
		\caption{\label{lm} The matter Lagrangian accumulation at different times. Note that there is no disturbance at the apparent horizon at late times. The disturbance is around the outgoing wave.}
\end{figure}

\clearpage

\begin{figure}[tbp]
		\begin{center}
			\includegraphics[scale=.38, draft=false, trim=1.5cm 1.5cm 2.5cm 2cm, clip=true]{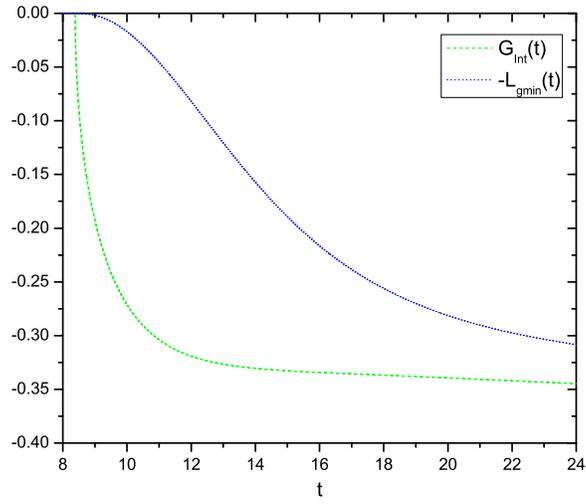}
		\end{center}
		\caption{\label{lgmin} The minimum in the Lagrangian accumulation and the interior Gibbs free energy $G_{int}$ as a function of time. The interior Gibbs free energy is given by $-r_{+}$.}
\end{figure}

\clearpage


\clearpage

\begin{figure}[tbp]
		\begin{center}
			\includegraphics[scale=.38, draft=false, trim=1.5cm 1.5cm 2.5cm 2cm, clip=true]{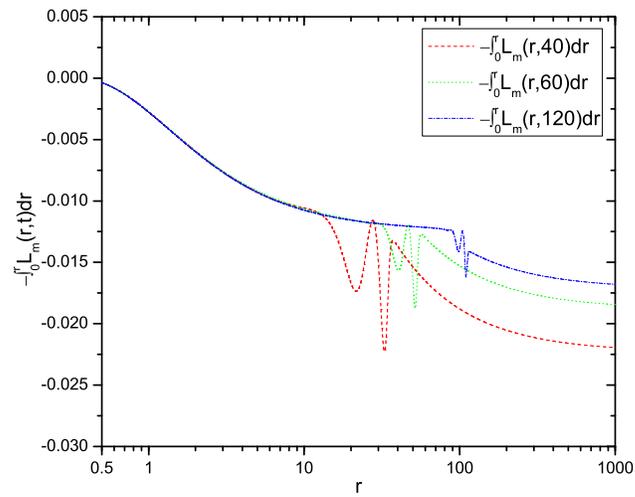}
		\end{center}
		\caption{\label{lmat2} The matter Lagrangian accumulation in the exterior plotted at different times.}
\end{figure}

\clearpage

\begin{figure}[tbp]
		\begin{center}
			\includegraphics[scale=.38, draft=false, trim=1.5cm 1.5cm 2.5cm 2cm, clip=true]{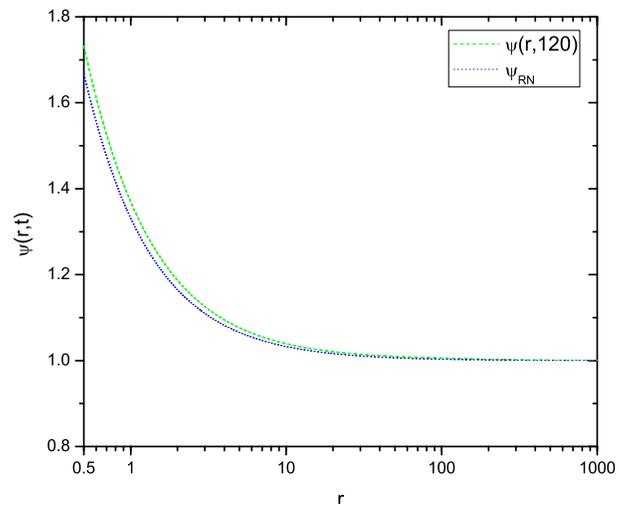}
		\end{center}
		\caption{\label{psiE} The conformal factor $\psi$ at $t=120$ is compared to its theoretical expectation in the exterior.}
\end{figure}

\clearpage

\section*{Acknowledgments}
A.E. acknowledges support from a discovery grant of the National
Science and Engineering Research Council of Canada (NSERC). H.B. 
acknowledges financial support from a Bishop's Senate Research Grant.

\end{document}